\newcommand{\zarticle}{z-article}

\newcommand{\useclass}{z-article}
\documentclass[letterpaper,10pt,twocolumn]{z-article}

\ifx\useclass\zarticle
\usepackage{S/oneinchmargins}  %
\usepackage{amsthm}
\usepackage[T1]{fontenc}
\usepackage{newtxtext,newtxmath}  %
\usepackage[square,numbers,comma,sort&compress]{natbib} %
\usepackage[breaklinks,hidelinks,colorlinks]{hyperref}
\usepackage{graphicx}
\usepackage[protrusion=true,expansion=true,nopatch=footnote]{microtype}
\usepackage{etoolbox}
\fi

\usepackage{multirow}
\usepackage{setspace}
\usepackage{relsize}
\usepackage{calc}
\usepackage[dvipsnames]{xcolor}
\usepackage{enumerate}
\usepackage{enumitem}
\setlist[itemize]{noitemsep,topsep=0pt,leftmargin=*}
\usepackage{booktabs} %
\usepackage[normalem]{ulem} %
\usepackage[ruled,vlined]{algorithm2e} %
\usepackage{float}
\usepackage{amsmath}
\usepackage{rotating}
\usepackage{xspace}
\usepackage{floatflt}
\usepackage{wrapfig}
\usepackage{alltt}
\usepackage{epstopdf}
\usepackage{subcaption}
\usepackage{listings}
\usepackage{fancyvrb}

\usepackage{colortbl} %
\usepackage{tcolorbox}
\tcbuselibrary{skins,breakable}
\usepackage{tabularx}

\usepackage{wasysym}  %
\usepackage{pifont}   %
\usepackage{upgreek}
\usepackage{circledsteps}

\usepackage{amssymb,amsfonts}
\usepackage{soul}

\usepackage{tikz, pgfplots, pgfplotstable}
\pgfplotsset{compat=1.16}
\usetikzlibrary{patterns}
\usepgfplotslibrary{fillbetween}
\usepackage{bm}

\usepackage{url}  %
\usepackage{xurl}
\usepackage{lipsum}

\makeatletter
\def\sectionautorefname#1{\S\@gobble#1}
\makeatother
\hypersetup{colorlinks=true,citecolor=teal,linkcolor=blue,urlcolor=blue}
\hypersetup{pdfauthor={},pdftitle={},pdfsubject={},pdfkeywords={}}
\begin{document}
\pagenumbering{arabic}
\definecolor{blue}{rgb}{0.38, 0.51, 0.71} %
\definecolor{darkblue}{RGB}{17, 42, 60} %
\definecolor{red}{RGB}{175, 49, 39} %

\definecolor{orange}{RGB}{217, 156, 55} %
\definecolor{green}{RGB}{144, 169, 84} %
\definecolor{palegreen}{RGB}{197, 184, 104} %

\definecolor{yellow}{RGB}{250, 199, 100} %
\definecolor{brokenwhite}{RGB}{218, 192, 166} %
\definecolor{brokengrey}{rgb}{0.77, 0.76, 0.82} %
\definecolor{fgray}{gray}{0.9}
\definecolor{epictint}{HTML}{E0F0EE} %

\makeatother

\def \ns {ns\xspace}
\def \us {$\upmu$s\xspace}  %
\def \usb {\mathbf{\upmu}\text{s}}

\def \xmark {\ding{55}} %
\def \cmark {\ding{51}} %
\newcommand{\hy}{\checkmark\kern-1.3ex\raisebox{.65ex}{\rotatebox[origin=c]{125}{\textbf{--}}}} %
\def \yes {\checkmark}
\def \no {\xmark}

\newcommand{\na}{-} %
\def \hi {\CIRCLE} %
\def \low {\Circle} %
\def \partial {\LEFTcircle} %

\def \cm {\checkmark}

\def \ra {$\rightarrow$}
\newcommand{\myra}[1]{\raisebox{-2pt}{$\xrightarrow{\texttt{#1}}$}}

\def \tms {$\times$\xspace}
\def \roughly {{\small $\sim$}}

\def \whitecircle {$\ocircle$}
\def \blackcircle {\ding{108}}
\def \whitesquare {$\Box$}
\def \blacksquare {\ding{110}}
\def \whitediamond {$\Diamond$}
\def \blackdiamond {\ding{117}}
\def \whitetriangle {$\bigtriangleup$}
\def \blacktriangle {\ding{115}}
\def \whitedtriangle {$\bigtriangledown$}
\def \blackdtriangle {\ding{116}}

\newcommand{\circled}[1]{\textbf{\raisebox{.5pt}{\textcircled{\raisebox{-.9pt}{#1}}}}}

\newcommand{\vtwenty}{\vspace{20pt}}
\newcommand{\vfifteen}{\vspace{15pt}}
\newcommand{\vten}{\vspace{10pt}}
\newcommand{\vfive}{\vspace{5pt}}
\newcommand{\vthree}{\vspace{3pt}}
\newcommand{\vtwo}{\vspace{2pt}}

\newcommand{\vminfive}{\vspace{-5pt}}
\newcommand{\vminten}{\vspace{-10pt}}
\newcommand{\vminfifteen}{\vspace{-15pt}}
\newcommand{\vmintwenty}{\vspace{-20pt}}

\def \hmina {\hspace{-0.1in}}
\def \hminb {\hspace{-0.2in}}

\newcommand{\ub}[1]{\underline{{\bf #1}}}
\newcommand{\ts}[1]{{\tt{\small#1}}}
\newcommand{\tts}[1]{{\tt{\footnotesize#1}}}

\newcommand{\bquote}{\vspace{-0.25cm} \begin{quote}}
\newcommand{\equote}{\end{quote}\vspace{-0.2cm} }
\def \sec {\S}

\def \nospace {
  \setlength{\itemsep}{0pt}
  \setlength{\parskip}{0pt}
  \setlength{\parsep}{0pt}
}

\newcommand{\myquote}[1]{
\begin{quote}
\centering
\small
\textit{#1}
\end{quote}
}

\newenvironment{enumerate2}{
\begin{enumerate}[leftmargin=*] \vminfive
  \setlength{\itemsep}{4pt}
  \setlength{\parskip}{0pt}
  \setlength{\parsep}{0pt}
}{
  \end{enumerate}
}

\newenvironment{itemize2}{
    \begin{itemize}[leftmargin=10pt] %
        \setlength{\itemsep}{2pt}
        \setlength{\parskip}{0pt}
        \setlength{\parsep}{0pt}
}{
  \end{itemize}
}

\newcommand{\rbt}[1]{\textcolor{red}{\textbf{#1}}}
\def \vvvnb {\vfifteen \noindent $\bullet$~}
\def \vvnb {\vten \noindent $\bullet$~}
\def \vnb {\vfive \noindent $\bullet$~}
\def \vn {\vfive \noindent}

\def \mb {\vspace{8pt}\nb}
\def \tb {\vspace{8pt}\nb}

\def \vvni {\vten \noindent}
\def \vni {\vfive \noindent}
\def \nb {\noindent $\bullet$~}
\def \ni {\noindent}
\def \bb {$\bullet$~}

\newcommand{\hypo}[1]{
    \begin{quote}
        \stepcounter{HYPO}{\bf Hypothesis \arabic{HYPO}:}
        {\em #1}
    \end{quote}
}

\newcommand{\taskformat}[2]{#1\textsc{#2}}

\newcommand{\task}[3]{
    \begin{quote}
    \phantomsection
    \hypertarget{task#1#2}{}
    {\bf Task \taskformat{#1}{#2}:}
    {\em #3}
    \end{quote}
}

\newcommand{\tasklink}[2]{\hyperlink{task#1#2}{\taskformat{#1}{#2}}}

\newcounter{HYPO}
\newcounter{TASK}

\newcommand{\rs}{{ResearchStaff$_1$}}
\newcommand{\pd}{{\bf Postdoc$_1$}}
\newcommand{\raOne}{{\bf RA$_1$}}
\newcommand{\raTwo}{{\bf RA$_2$}}
\newcommand{\ndv}{{\bf NDV}}
\newcommand{\ug}{{\bf Undergrad$_1$}}

\newcommand{\sssubsection}[1]{\vten\ni\textbf{\large{\textsc{#1}}}}

\newcounter{mysubcounter}
\setcounter{mysubcounter}{1}
\newcommand{\mysub}[1]{\vtwo\ni\textcolor{black}{\textbf{\textbf{#1}.\stepcounter{mysubcounter}}}}

\newcommand{\emptypage}{
\newpage
(empty page)
}

\newcommand{\myrotate}[1]{\begin{rotate}{90} {\bf #1} \end{rotate}}

\newcommand{\mycaption}[3]{
    \begin{spacing}{0.95}
        \caption{
            \label{#1}
            {\small\bf #2. }
            {\normalfont\em\small #3}
        }
    \end{spacing}
}

\newcommand{\eg}{\textit{e.g.}}
\newcommand{\ie}{\textit{i.e.}}
\newcommand{\etal}{\textit{et al.}}
\newcommand{\etc}{etc.}
\def \th {$^{th}$\xspace}

\newcommand{\sstar}{$^{*}$}
\newcommand{\stwostars}{$^{**}$}
\newcommand{\stristars}{$^{***}$}
\newcommand{\srealstar}{$^{\star}$}
\newcommand{\sdag}{$^{\dag}$}
\newcommand{\sddag}{$^{\ddag}$}

\newcounter{Xcounter}
\newcommand{\xxxreset}{\setcounter{Xcounter}{1}}
\newcommand{\xxx}{{\footnotesize\textcolor{red}{\textbf{xxx$_{\arabic{Xcounter}}$}\stepcounter{Xcounter}}~}}

\newcommand{\xxxinfig}{\textcolor{red}{\textbf{xx}}} %

\newcounter{Fcounter}
\setcounter{Fcounter}{1}

\newcommand{\myfinding}[1]{\vtwo\noindent{{\rbt{Finding \#\arabic{Fcounter}:} \stepcounter{Fcounter}}#1}}

\newtcolorbox[auto counter, number within=section]{hlbox}[2][]{
  colback=gray!25, %
  colframe=gray!0,      %
  fonttitle=\bfseries,  %
  coltitle=black,       %
  boxrule=.5pt,       %
  left=1pt,             %
  right=1pt,            %
  top=.5pt,              %
  bottom=1.5pt,           %
  boxsep=.5pt,           %
  arc=4pt,              %
  before skip=1pt,      %
  after skip=1pt,       %
}
\newtcolorbox[auto counter, number within=section]{findingbox}[2][]{
    colback=blue!8,
    colframe=gray!90,
    fonttitle=\bfseries,
    coltitle=black,
    boxrule=.5pt,
    left=2pt, right=2pt, top=2pt, bottom=2pt,
    boxsep=1pt, arc=0pt,
    before skip=2pt, after skip=2pt,
    breakable
}
\newcommand{\moatfinding}[1]{
\begin{findingbox}{}
    \noindent{{\small\bf{{Takeaway \#\arabic{Fcounter}:}} \stepcounter{Fcounter}}#1}
\end{findingbox}
}

\def \opone {\texttt{op1}\xspace}
\def \optwo {\texttt{op2}\xspace}
\def \matA {\texttt{\%mA}\xspace}
\def \matB {\texttt{\%mB}\xspace}
\def \transmatA {\texttt{\%tmA}\xspace}
\def \gemmt {\texttt{matmult-t1}\xspace}
\def \nfpmap {\texttt{@nfp}\xspace}
\def \armmap {\texttt{@arm}\xspace}
\def \npumap {\texttt{@npu}\xspace}
\def \anymap {\texttt{@any}\xspace}

\def \entrynode {\texttt{entry}\xspace}
\def \exitnode {\texttt{exit}\xspace}
\def \tlocal {\texttt{task-local}\xspace}
\def \tshared {\texttt{task-shared}\xspace}
\def \persist {\texttt{persistent}\xspace}

\def \muop {$\upmu$ops\xspace}
\newcommand{\hwt}[1]{\texttt{#1}}
\newcommand{\loctype}[1]{\texttt{@#1}}
\newcommand{\csdcmd}[1]{\textit{#1}}
\def \sdfg {{SDG}\xspace}
\def \tdfg {{TDG}\xspace}
\def \figx {\rbt{Figure ?}\xspace}
\newcommand \myautoref[2]{\hyperref[#1]{\autoref*{#1}#2}}
\def \sys {{\textsc{Epic}}\xspace}
\def \mydsl {\textsc{\sys-dsl}\xspace}
\def \mycc {\textsc{Epic-cc}\xspace}
\def \myrt {\textsc{Epic-rt}\xspace}

\newlength{\outerradius}
\newlength{\innerradius}
\setlength{\outerradius}{6pt}
\setlength{\innerradius}{4pt}

\newcommand{\progresscircle}[1]{
  \begin{tikzpicture}
    \fill[black!80] (0,0) circle (\outerradius);
    \fill[gray!70,draw=white,line width=1pt] (0,0) -- (0, \outerradius+0.5pt)
      arc (90:90-3.6*#1:\outerradius+0.5pt) -- (0,0);
    \fill[white] (0,0) circle (\innerradius);
  \end{tikzpicture}
}

\newcommand{\mycodebox}[2][]{%
  \begin{tcolorbox}[
    enhanced,
    breakable,
    colback=white!0,      %
    colframe=white!0,     %
    boxrule=0pt,
    overlay={%
      \begin{scope}[shift={(frame.north west)}]
          \foreach \y/\h/\c in {#1} {%
            \fill[\c, rounded corners] (0,-\y\baselineskip) rectangle ++(\linewidth,-\h\baselineskip);
          }
      \end{scope}
    }
  ]
  #2%
\end{tcolorbox}}

\newcommand{\codequote}[2][]{
    \node at (#1) {
    \ts{\textcolor{red}{#2}}
    };
}

\newenvironment{quotedcode}[1]{
    \begin{tikzpicture}[
        every node/.style={anchor=north west,inner sep=0pt},
        x=\baselineskip, y=\baselineskip,
      ]
        \node (code) at (0,0) {#1};
}
{\end{tikzpicture}}

\newcommand{\cEpic}{\ts{Epic}\xspace}
\newcommand{\cHost}{\ts{Host}\xspace}
\newcommand{\cSync}{\ts{Sync}\xspace}
\newcommand{\cStateless}{\ts{Stateless}\xspace}
\newcommand{\cPageCache}{\ts{Page Cache}\xspace}
\newcommand{\cSingleTask}{\ts{Single-Task}\xspace}
\newcommand{\cSharedFIFO}{\ts{Shared FIFO}\xspace}
\newcommand{\cNoFusion}{\ts{NoFusion}\xspace}
\newcommand{\cNoPipelining}{\ts{NoPipelining}\xspace}
\newcommand{\cNFPOnly}{\ts{NFP-Only}\xspace}
\newcommand{\cNPUOnly}{\ts{NPU-Only}\xspace}
\newcommand{\cARMPrefetch}{\ts{ARM + Prefetch}\xspace}
\newcommand{\cARM}{\ts{ARM}\xspace}
\newcommand{\cNFP}{\ts{NFP}\xspace}
\newcommand{\cLambdaIO}{\ts{$\lambda$-IO}\xspace}
\newcommand{\cBiscuit}{\ts{Biscuit}\xspace}
\newcommand{\cAssasin}{\ts{Assasin}\xspace}
\newcommand{\cDSCS}{\ts{DSCS}\xspace}
\newcommand{\cInsider}{\ts{Insider}\xspace}
\newcommand{\cStageOne}{\ts{Stage 1}\xspace}
\newcommand{\cStageTwo}{\ts{Stage 2}\xspace}
\newcommand{\cHostTransfer}{\ts{Host Transfer}\xspace}
\newcommand{\wKVGet}{\ts{KV Get}\xspace}
\newcommand{\wJoin}{\ts{Join}\xspace}
\newcommand{\wDLRM}{\ts{DLRM}\xspace}
\newcommand{\wVA}{\ts{VA}\xspace}
\newcommand{\wEC}{\ts{EC}\xspace}
\newcommand{\wHR}{\ts{HR}\xspace}
\newcommand{\wMQA}{\ts{MQA}\xspace}
\newcommand{\wStatsThirtyTwo}{\ts{Stats32}\xspace}
\newcommand{\wStatsSixtyFour}{\ts{Stats64}\xspace}
\newcommand{\wKNN}{\ts{KNN}\xspace}
\newcommand{\wGrep}{\ts{Grep}\xspace}
\newcommand{\wBitmap}{\ts{Bitmap}\xspace}

\RestyleAlgo{ruled} %
\LinesNumbered %
\SetAlgoVlined %
\DontPrintSemicolon %
\SetAlFnt{\footnotesize} %
\SetInd{0.5em}{0.5em}
\renewcommand{\CommentSty}[1]{\textnormal{\ttfamily\textcolor{red}{#1}}\unskip}
\SetKwComment{Comment}{$\triangleright$\ }{} %

\newcommand{\coloredkw}[1]{\textcolor{black}{\textbf{#1}}}
\SetKwFor{ForEach}{\coloredkw{foreach}}{:}{}
\SetKwFor{For}{\coloredkw{for}}{:}{}
\SetKwIF{If}{ElseIf}{Else}{\coloredkw{if}}{:}{\coloredkw{else if}}{else:}{}
\SetKwProg{Fn}{Function}{:}{}
\SetKw{KwContinue}{continue}

\def \examplealgo {
\begin{algorithm}
\caption{Quick Sort}
\SetKwFunction{FQuickSort}{QUICKSORT}
\SetKwFunction{FPartition}{PARTITION}
\Fn{\FQuickSort{$A, low, high$}}{
    \If{$low < high$}{
        $pivotIndex \gets \FPartition(A, low, high)$\;
        \FQuickSort{$A, low, pivotIndex - 1$}\tcp*[r]{inlin}
        \FQuickSort{$A, pivotIndex + 1, high$}\Comment*[r]{inline}
        \tcc{This is another comment.}
        \Comment{This is another comment.}
    }
}
\Fn{\FPartition{$A, low, high$}}{
    $pivot \gets A[high]$\;
    $i \gets low - 1$\;
    \For{$j \gets low$ \KwTo $high - 1$}{
        \If{$A[j] \leq pivot$}{
            $i \gets i + 1$\;
            Swap $A[i]$ and $A[j]$\;
        }
    }
    Swap $A[i + 1]$ and $A[high]$\;
    \KwRet{$i + 1$}\;
}
\end{algorithm}}

\def \mytitle {{Programming In-Storage Computing with Located, Stateful Dataflow}}
\author{%
  Yuyue Wang\\{\small\texttt{yuyue@cs.ucla.edu}}
  \and Zhenyu Zhang\\{\small\texttt{ashfish.zzy@gmail.com}}
  \and Glenn Reinman\\{\small\texttt{reinman@cs.ucla.edu}}
  \and Huaicheng Li\\{\small\texttt{huaicheng@cs.vt.edu}}%
}
\date{}
\hypersetup{
  pdftitle={Programming In-Storage Computing with Located, Stateful Dataflow},
  pdfauthor={Yuyue Wang; Zhenyu Zhang; Glenn Reinman; Huaicheng Li}
}
\title{\mytitle}

\maketitle
\thispagestyle{plain}
\pagestyle{plain}
\begin{abstract}
In-storage computing (ISC) reduces host--storage data movement by
executing computation inside computational storage devices (CSDs).
For multi-stage applications, realizing these benefits requires
coordinating data placement, I/O--compute overlap, and device-resident
state across the workflow, yet existing interfaces lack a unified
abstraction for these decisions.
We present \sys, an NVMe-based ISC stack that provides this abstraction
by capturing data residency and lifetime in the program:
location types declare logical residency,
dataflow derives lifetimes for intermediate values and operation state
within an invocation, and a keep primitive extends selected state across
invocations.
These semantics expose the complete offloaded workflow as a located,
stateful dataflow.
A storage-aware compiler transforms this workflow, performs
movement-aware logical mapping and fusion, and exposes I/O--compute
overlap; a runtime completes the plan using execution-time information,
asynchronously binding work to physical resources and managing device-resident
state.
Across 12 file-scanning, database, and machine learning workloads, \sys is
1.6$\times$ faster on average than the strongest of five prior ISC
systems, while achieving 4.2$\times$ speedup on average
and up to 16.1$\times$ over the corresponding host baselines, and
reducing application-side code by up to 14$\times$ in
our implementations.
\ifx\useclass\zarticle\else\vminten\fi  %
\end{abstract}

\setcounter{page}{1}

\section{Introduction}
\label{s:intro}

\providecommand{\citeall}{\cite{nascent.fpga21,
bluedbm.isca15, grafboost.isca18, extrav.vldb17, rmssd.hpca22,
recssd.asplos21, cognitivessd.atc19, holistic-gnn.fast22, glist.atc21,
smart-inf.hpca24, genstore.asplos22, inspire.isca22,
ecssd.isca23, optimstore.hpca23, deepstore.micro19, beacongnn.hpca24,
instinfer.hpca25, behemoth.fast21, yoursql.vldb16, smartsage.isca22,
dockerssd.hpca24, flashabacus.eurosys18, fpim3d.micro22,
crossbit.micro25, sting.dac24, iskeva.lctes22, ibex.vldb14}}

In-storage computing (\textbf{ISC}) processes data within the storage
device and returns compact results to the host, reducing the data
movement that often dominates data-intensive
computing~\cite{trainbottleneck.sigmod22,activeflash.fast13}. Over two decades, ISC has
evolved from early ActiveDisk prototypes~\cite{activedisk.asplos98} to
commercial computational storage devices (\textbf{CSDs})~\cite{smartssd-gen2.fms22,
scaleflux-csd.web,ibm-flashsystem.web} and increasingly
complex applications~\cite{deepstore.micro19,polardb.fast20,genstore.asplos22,
ecssd.isca23,megis.isca24,smartanns.atc24,salientstore.pact25}.
CSDs are also becoming heterogeneous, incorporating processing units
(\textbf{PUs}) such as general-purpose ARM cores, neural processing units
(\textbf{NPUs}), and near-flash processors (\textbf{NFPs}) for different
classes of computation~\cite{lambda-io.fast23,assasin.micro22,dscs.asplos24}. Meanwhile,
the NVMe Computational Programs and Subsystem Local Memory
(\textbf{SLM}) specifications standardize mechanisms for executing device
programs and accessing device-local memory~\cite{nvmecsdspec.web,
nvmeslmspec.web}. The hardware and standardized execution mechanisms
have arrived; what remains missing is a workflow-level software
abstraction for expressing and optimizing multi-stage computation across
storage, device-local state, and heterogeneous compute.

Existing ISC systems expose useful pieces of the device execution path
through transparent offloading, customized interfaces, and programmable
device frameworks~\cite{willow.osdi14,insider.atc19,metalfs.eurosys20,
lambda-io.fast23,activepy.dac23,biscuit.isca16}.
\sec\ref{s:rel} compares these programming and execution models in detail.
These mechanisms improve individual aspects of offloading and I/O, but
generally expose computation as isolated offload units or within
system-specific scopes, leaving placement, movement, overlap, and
device-resident state to separate mechanisms or fixed design choices.
Such fragmented designs do not provide the coordination needed by
heterogeneous CSDs whose processors occupy different points of the storage
data path, nor by multi-stage, stateful applications in which placement,
I/O--compute overlap, and device-resident state interact
(\sec\ref{s:motiv}).

Our insight is to optimize the complete offloaded workflow: computation,
flash access, host transfer, intermediate values, and state reused across
invocations form one \emph{located, stateful dataflow}. A storage-facing
contract captures \emph{where values reside} and \emph{how long they live}.
Together with program dependencies and shapes, this representation enables
three coordinated decision families: \emph{spatial} decisions partition,
map, and fuse computation across processing resources; \emph{temporal}
decisions overlap I/O and computation within and across kernels; and
\emph{stateful} decisions retain device state across storage accesses,
dependent tasks, and invocations.

We present \textbf{\sys}, an NVMe-based ISC stack that realizes this
model. Developers express offloaded logic in a Python eDSL whose
function interfaces use location types to declare logical storage residency,
including \loctype{host}, \loctype{slm}, and \loctype{flash}.
Within an invocation, \sys derives the lifetimes of intermediate values
and operation state from producer--consumer dataflow; state that must
outlive the invocation is kept explicitly with the \ts{epic.keep}
primitive.
From this program, \sys's compiler (\mycc) automatically derives a movement-aware
logical execution plan under a cost model, including logical PU mapping,
fusion, tiling, and computation--I/O pipelining.
Its runtime (\myrt) asynchronously realizes this logical plan, resolving
flash-translation-layer (\textbf{FTL}) managed placement, binding ready
kernels to physical processor instances and flash channels, and managing
device-resident state, including synchronizing conflicting accesses to
mutable kept objects.
To our knowledge, \sys is the first ISC system to automatically derive a
coordinated workflow plan for heterogeneous PU placement,
cross-stage fusion, I/O--compute overlap, and device-resident state
from a single storage-typed program.

We implement \sys on a reference heterogeneous CSD combining embedded
cores, tensor acceleration, and per-channel programmable compute, and
evaluate it using FEMU for full-system storage-path behavior and
cycle-accurate component simulators for accelerator timing. Across 12
workloads spanning file scanning, database queries, and end-to-end
machine learning (ML) pipelines, \sys accelerates every workload over its
respective host
baseline, delivering 4.2$\times$ speedup on average and up to
16.1$\times$ on DLRM embedding lookups. Compared with five state-of-the-art
ISC systems ($\lambda$-IO, Biscuit, Assasin, DSCS, and
Insider~\cite{lambda-io.fast23,biscuit.isca16,assasin.micro22,dscs.asplos24,insider.atc19}),
\sys is 1.6$\times$ faster on average than the strongest system.
It also reduces application-side code by up to 14$\times$ in
our implementations.

In summary, we make the following contributions:

\begin{itemize}[leftmargin=*, topsep=2pt, itemsep=1pt]

\item We define a CSD-specific workflow contract in which \loctype{host},
\loctype{slm}, and \loctype{flash} describe storage residency rather than
processing units, while dataflow-derived lifetimes and \ts{epic.keep}
expose how long device-resident values must live. From this single
storage-typed program, \mycc automatically derives heterogeneous logical PU
placement, fusion, and I/O--compute overlap over the workflow.

\item We design a coordinated compiler/runtime stack that combines
movement-aware workflow transformation and logical mapping with
asynchronous execution, FTL-aware physical dispatch, and device-resident
state management.

\item We implement \sys as a full-stack ISC system on a reference
heterogeneous CSD and evaluate it across 12 workloads. Its automatically
generated workflow plans are 1.6$\times$ faster on average than the
strongest prior system, while
achieving 4.2$\times$ speedup on average over host
baselines and reducing application-side code by up to 14$\times$ in our
implementations.

\end{itemize}

\section{Background and Motivation}
\label{s:bg}

\subsection{CSD Architecture and Interfaces}
\label{s:bg:csd}

A CSD extends a conventional SSD
with device-side processing while retaining the underlying storage path.
Host requests cross the PCIe/NVMe interface to the SSD controller
(\autoref{fig:hw}), whose
FTL maps logical addresses onto physical
flash pages distributed across channels and chips. Device DRAM supports the
storage control plane and application-visible device-local memory. Modern
CSDs increasingly place multiple PUs along this
path: embedded ARM cores provide general-purpose execution and control,
NPUs accelerate dense tensor operations
over data staged through shared device DRAM, and per-channel near-flash
processors exploit channel locality for streaming
computation~\cite{lambda-io.fast23,assasin.micro22,dscs.asplos24,
smartssd-gen2.fms22,dockerssd.hpca24}. These resources differ in compute
capability, locality, and access to shared memory, so execution placement is
part of the storage data path rather than an isolated accelerator decision.
We use this organization as the reference CSD for both system design and
evaluation.

\begin{figure}[t!]
\centering
\includegraphics[width=\columnwidth]{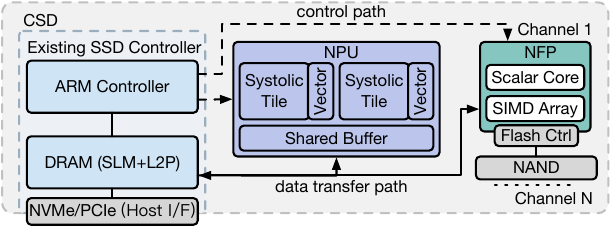}
\vminten
\mycaption{fig:hw}{Reference CSD organization used by \sys}{ARM controller,
NPU, and per-channel NFPs share device DRAM and are coordinated behind the
host NVMe interface.}
\end{figure}

\textbf{NVMe execution interface.}
NVMe provides a request/completion interface between host software and the
storage device. The Computational Programs and Subsystem Local Memory
specifications extend this interface with standardized mechanisms for
invoking device-resident programs and accessing
SLM~\cite{nvmecsdspec.web,nvmeslmspec.web}. They define how
computation is submitted and how device-local memory is exposed, but not how
a multi-stage application workflow should be partitioned, fused, or
scheduled across storage and compute resources. SLM is volatile device-local storage;
\sys uses it for reuse during device operation such as across invocations.

\subsection{Workflow Coordination}
\label{s:motiv}

Prior ISC systems address individual pieces of this problem, but do not
provide a common workflow representation that coordinates placement,
I/O--compute overlap, and device-resident state across the application;
\sec\ref{s:rel} compares their programming and execution models.

\begin{figure}[t]
\centering
\captionsetup[subfigure]{skip=2pt}
\includegraphics[width=\columnwidth]
    {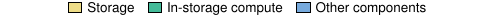}\par
\vspace{-3pt}

\begin{subfigure}[b]{0.495\columnwidth}
    \centering
    \includegraphics[width=\linewidth]
        {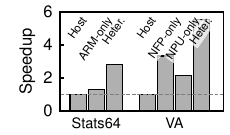}
    \caption{Speedup over host.}
    \label{fig:motivation-speedup}
\end{subfigure}
\hfill
\begin{subfigure}[b]{0.495\columnwidth}
    \centering
    \includegraphics[width=\linewidth]
        {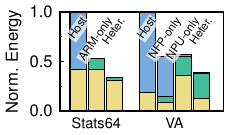}
    \caption{Energy breakdown.}
    \label{fig:motivation-energy}
\end{subfigure}

\vminten
\mycaption{fig:motivation}{Placement affects both CSD performance and energy}{%
Speedup and energy for \wStatsSixtyFour and \wVA under the same
placements, normalized to the host baseline.
\ts{Heter.} combines PU classes rather than fixing execution to one class.
Energy is decomposed into storage (flash and device DRAM), in-storage
compute, and other components (host-side compute and memory, including
accelerators, plus PCIe transfers).
Energy: host CPU measured with RAPL~\cite{rapl.tompecs18}; the rest
estimated from specifications, models, and synthesis~\cite{amber.micro18,
cortex-r82.web,regate.micro25,drampower.web,pcie-phy.asscc18}.}
\end{figure}

The need for coordination follows from the CSD substrate. Storage parallelism
is distributed across flash channels, while data movement converges through
shared device DRAM and the host PCIe link. Compute resources follow a
different hierarchy: lightweight per-channel engines exploit locality and
bandwidth, whereas more capable shared processors favor compute-intensive
kernels. Placement must therefore account for both data location and compute
demand. \autoref{fig:motivation-speedup} illustrates the benefit of
heterogeneous placement: it is 2.2$\times$ faster than ARM-only execution
on \wStatsSixtyFour and 1.7$\times$ faster than NFP-only execution on
\wVA.
Placement also affects energy:
\autoref{fig:motivation-energy} shows that in-storage execution reduces
host-side compute and data-transfer energy.
On \wVA, either NFP-only or NPU-only execution consumes roughly 40\%
more energy than heterogeneous placement.
These results motivate joint workflow planning for placement and data
movement.

Flash access latency, orders of magnitude above DRAM's, additionally makes
I/O--compute overlap essential. Storage and AI pipelines also reuse
intermediates, model state, or caches across invocations; returning such
state through the host discards locality and reuse.
CSD software must therefore coordinate spatial placement, temporal
I/O--compute overlap, and stateful device-resident reuse.

These demands are coupled within a workflow. Consider a flash-resident
filter followed by computation over the selected rows: keeping the stages
separate materializes the selected data in SLM, whereas fusion can consume
those rows directly along the near-flash path. That choice changes the
intermediate's location, eligible PU, data movement, tiling and buffer
lifetime, and which I/O can overlap with computation. Likewise, an
application-managed embedding cache should remain in SLM across lookup
invocations rather than be reconstructed or round-tripped through the host.

These decisions also depend on information available at different times.
Logical residency, tensor shapes, and workflow dependencies enable
compile-time workflow transformation and logical PU selection. FTL
translation and PU availability are dynamic. The compiler therefore
produces a logical execution plan, while the runtime schedules ready work
and binds it to physical resources using execution-time information.

\section{\sys Design}
\label{s:des}

\sys's compiler (\mycc) and runtime (\myrt) coordinate placement,
I/O--compute overlap, and device-resident state through the workflow
abstraction described below. \autoref{tab:families} summarizes their
responsibilities. Stateful decisions cover lifetime, reuse, and
consistency of device-resident state.

\subsection{Programming Model}
\label{s:des:dsl}

\sys exposes the offloaded workflow through \mydsl, a Python-embedded
domain-specific language in which the program provides the storage-facing
semantics needed by the rest of the stack. In \sys, a workflow is the
dependency graph of one compiled offloaded function together with
references to device-resident objects whose lifetimes may span computation
stages and function invocations. Developers write the computation in a
sequential, declarative style; \sys combines declared residency, dataflow-derived
lifetimes, the explicit \ts{epic.keep} primitive, and program dependencies
to reconstruct the workflow optimized and executed by later stages.

\begin{table}[t!]
\centering
\scriptsize
\setlength{\tabcolsep}{3pt}
\renewcommand{\arraystretch}{1.08}
\begin{tabular}{@{}p{0.17\columnwidth}p{0.39\columnwidth}p{0.39\columnwidth}@{}}
\textbf{Family} &
\textbf{Compile-time plan} &
\textbf{Runtime realization} \\
\hline

\textbf{Spatial} &
Logical PU mapping, fusion, and data-parallel partitioning
(\sec\ref{s:des:cc:spatial}) &
PU-instance binding and FTL-resolved channel dispatch
(\sec\ref{s:des:rt:ftl}) \\

\textbf{Temporal} &
Execution-unit formation, tiling, buffering, and software-pipelined
overlap (\sec\ref{s:des:cc:temporal}) &
Asynchronous dispatch, two-level scheduling, and resumable
execution (\sec\ref{s:des:rt:sched}) \\

\textbf{Stateful} &
Invocation-local lifetime derivation, \ts{epic.keep} metadata, and
synchronization points (\sec\ref{s:des:dsl}, \sec\ref{s:des:state}) &
Invocation-local state reclamation by reference counting;
cross-invocation state management and per-object critical-section
synchronization (\sec\ref{s:des:state}) \\

\end{tabular}
\vminfive
\mycaption{tab:families}{Decision families across binding times}{%
\sys coordinates three decision families across compile time and runtime.}
\vminfive
\end{table}

\begin{figure}[t!]
\hspace*{-0pt}\includegraphics[width=1.\columnwidth]{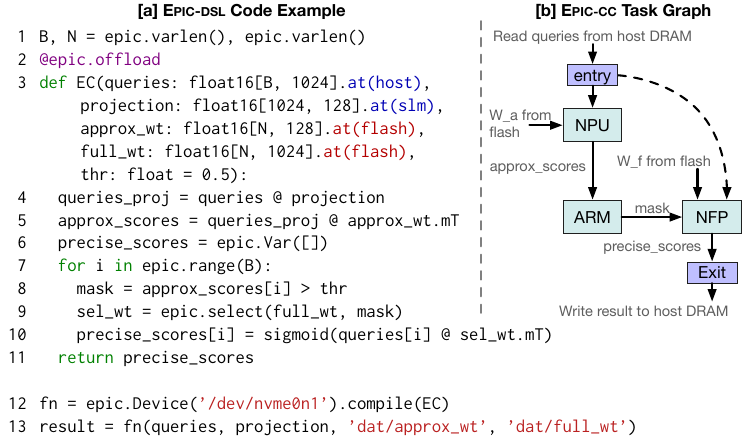}
\vmintwenty
\mycaption{fig:dsl-example}{From a location-typed function to an executable workflow}{%
(a)~\mydsl code for extreme classification.
(b)~The derived workflow graph preserves data locations and dependencies.}
\end{figure}

\subsubsection{The \mydsl}

An \sys offloaded function has parameters that carry ordinary type
information---scalars or typed tensors with explicit element precision
(\eg, FP32 or FP64)---together with data location types
(\loctype{host}, \loctype{slm}, or \loctype{flash}).
These locations describe the logical residency of inputs: host memory,
device-local SLM, or flash.

Since input residency affects storage accesses and data transfers, explicit
location types let \sys plan these operations ahead of time.
Alternatively, \sys also supports just-in-time compilation that infers
residency from runtime arguments, allowing location annotations to be omitted.
\autoref{fig:dsl-example}a (lines 1--11) illustrates the EC interface.

Inside the function, \mydsl represents tensor operations together with
structured control flow such as loops, branches, and returns, allowing
multi-stage pipelines to be expressed as one logical computation.
Developers may also register custom C functions as opaque operators via
\ts{@epic.extern}. Because locations are attached to values rather than
individual offload calls, the resulting dataflow makes movement between
host, SLM, and flash explicit to the system without requiring the
application to issue those transfers directly.

Within an invocation, \mycc derives the lifetimes of intermediate values
and operation state from producer--consumer dependencies in the workflow
graph. We call this invocation-local state \emph{CallState}.
To retain a value across invocations, the program uses the \ts{epic.keep}
primitive and returns its handle for subsequent invocations to reuse.
These retained objects form \emph{SharedState} and remain allocated until
the host explicitly releases them.
A parameter may additionally be marked \texttt{mut} to denote updatable
state; accesses to the same mutable object are mutually exclusive, while
independent objects may be accessed concurrently.
\sec\ref{s:des:state} defines the corresponding lifetime, reclamation,
and synchronization semantics.

From the host application's perspective, an offloaded function is exposed
through a compiled device handle. As shown in
\autoref{fig:dsl-example}a (lines 12--13), the application compiles the
function for a CSD with \ts{Device(...).compile()} and invokes the
resulting handle as a normal callable. The host--CSD request path,
argument staging, and asynchronous NVMe execution behind this interface
are described in \sec\ref{s:des:rt:nvme}.

\subsubsection{Workflow Construction}

During compilation, \mycc transforms the function and its interface
metadata into a workflow graph, as shown in \autoref{fig:dsl-example}b.
The graph preserves data and control dependencies together with
each value's logical storage location. In the EC example,
host-resident queries, SLM-resident
projection data, and flash-resident weight matrices become connected
stages of one workflow rather than independent offload requests. The
graph identifies required data movement and which dependent
operations may benefit from co-location, fusion, or overlap.

\subsection{Storage-Aware Compilation}
\label{s:des:cc}

The \mycc compiler turns the residency- and lifetime-aware workflow from
\sec\ref{s:des:dsl} into a logical execution plan for the CSD. It is
parameterized by a target device model describing available PU classes
and instance counts, SLM capacity and bandwidth, and flash-path
characteristics. Spatial decisions determine logical PU placement,
fusion, and parallel partitioning, while temporal decisions overlap
flash I/O with computation.

\subsubsection{Workflow IR and Cost Model}
\label{s:des:cc:ir}

The workflow graph constructed from \mydsl is lowered into a canonical
SSA-based intermediate representation with inferred types, shapes, and
constants. Within this IR, \mycc represents data and control dependencies
as an SSA-level dependency graph (\sdfg): nodes are operations, while
edges carry the logical residency information derived from the program.
The \sdfg serves as the planning
representation for the cost model and workflow transformations below.

\mycc evaluates alternative workflow plans with a resource-constrained
makespan model. For each operation, the model estimates flash-access
($T_{\mathrm{flash}}$), SLM-movement ($T_{\mathrm{slm}}$), and compute
($T_{\mathrm{compute}}$) time. It then schedules these activities on
flash, SLM, and per-PU-instance resource timelines, accounting for
contention and the number of available instances; the completion time of
the resulting schedule is the estimated plan cost. A transfer spanning
$n$ movement hops and $B$ bytes incurs
$nB/\mathrm{BW}_{\mathrm{slm}} + nT_{\mathrm{hop}}$ and reserves the
SLM timeline, where $\mathrm{BW}_{\mathrm{slm}}$ is the SLM bandwidth
and $T_{\mathrm{hop}}$ the fixed per-hop cost.

\begingroup
\postdisplaypenalty=10000
The model also captures overlap already represented in the planning IR.
For a pipelined tile-loop region, its steady-state stage cost is bounded
by
\begin{equation*}
\max\!\left(T_{\mathrm{flash}},\,
           T_{\mathrm{slm}},\,
           T_{\mathrm{compute}}\right),
\end{equation*}
rather than their sum. Target-specific double buffering and NFP
page-transfer/computation pipelining are added during lowering
(\sec\ref{s:des:cc:temporal}) and are not fed back into logical
mapping.
\par
\endgroup

Semantics-preserving canonical rewrites are applied to a fixed point
without cost gating. Logical device mapping uses a makespan-driven local
search over capability-compatible PU assignments. Fusion is
cost-directed: \mycc tentatively applies a fusion candidate, recomputes
the mapping induced by the transformed graph, and retains the fusion if
the resulting makespan does not increase; otherwise it restores the
previous graph. We evaluate the cost model's accuracy in
\sec\ref{s:eval}.

\subsubsection{Placement and Fusion}
\label{s:des:cc:spatial}

\begin{figure}[t]
\centering
\includegraphics[width=\columnwidth]{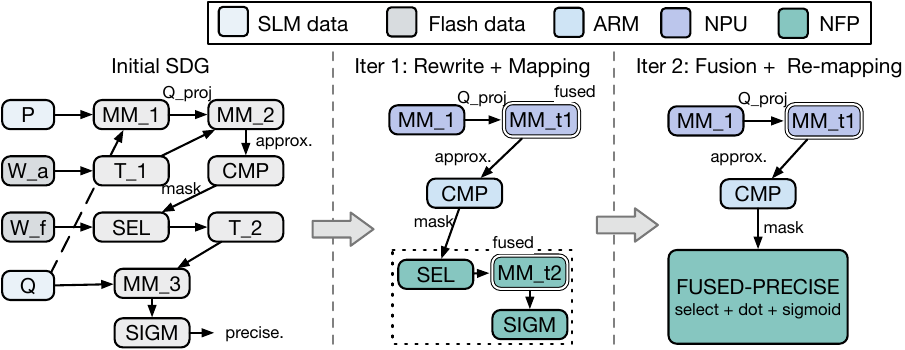}
\vminfive
\mycaption{fig:cc-opt}{Joint workflow transformation and logical mapping}{%
\sdfg evolution for the extreme-classification workload. Canonical
rewrites reduce the initial graph from eight to six operators. Two
fusion candidates are then accepted under their induced mappings,
reducing the graph to four operators.
Logical PU mappings change across transformation iterations.}
\end{figure}

\mycc first canonicalizes the workflow with semantics-preserving graph
rewrites that remove unnecessary intermediates. In the EC example
(\autoref{fig:dsl-example}a, \autoref{fig:cc-opt}), the
transpose--matmul pairs
\texttt{(T\_1,\,MM\_2)} and \texttt{(T\_2,\,MM\_3)}
are rewritten into transpose-aware matmul nodes
\texttt{MM\_t1} and \texttt{MM\_t2}, reducing the graph from eight to
six operators.

For logical mapping, capability rules first restrict and prioritize
candidate PU classes. Dense tensor operators such as matmul and
convolution preferentially target the NPU; element-wise, reduction,
gather, and select operations over \loctype{flash}-resident operands
target the NFP so that they can execute along the near-flash path;
opaque/external or precision-constrained computation targets ARM.
Control-flow and layout operations inherit a compatible neighboring
placement. These rules determine the feasible candidates rather than the
final assignment: \mycc uses the makespan model to search among them,
including the data movement induced at placement boundaries.

The EC workflow illustrates the resulting placement. After
canonicalization, \texttt{MM\_1} performs the dense query projection and
\texttt{MM\_t1} computes approximate scores; both map to the NPU. The
comparison operation consumes these NPU-produced scores and maps to ARM,
while flash-resident candidate selection, the precise matvec, and the
subsequent sigmoid are fused into an NFP operation that selects the required
rows directly from flash and computes their precise scores without materializing the selected
weight set in SLM. Two accepted fusion candidates reduce the six-operator
canonical graph to four operators. \mycc then considers a third fusion candidate, but the
resulting graph induces a different mapping whose estimated makespan is
higher, so the compiler rolls the transformation back and retains the
four-operator plan (\autoref{fig:cc-opt}). Thus capability, residency,
fusion, and movement cost jointly determine the logical ARM/NPU/NFP
placement.

Spatial planning also extracts data parallelism. When an operation is
applied independently across slices of a
\texttt{range}/\texttt{grid}/\texttt{tiles} iteration,
\mycc partitions it into kernels over disjoint index ranges. These
kernels retain the same logical PU class but can later execute on
different physical instances of that class, including multiple ARM cores
or the per-channel NFPs selected from the physical location of their
input pages (\sec\ref{s:des:rt:sched}, \sec\ref{s:des:rt:ftl}).

\subsubsection{Tiling and I/O Overlap}
\label{s:des:cc:temporal}

\begin{figure}[t]
\centering
\includegraphics[width=.92\columnwidth]{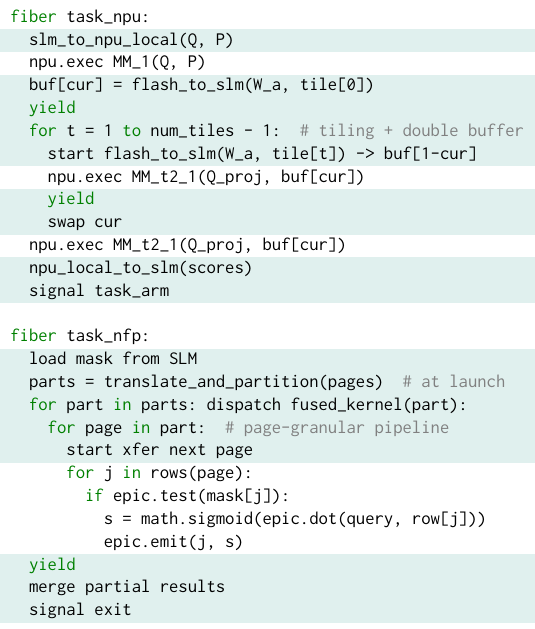}
\vminten
\mycaption{fig:fiber-code}{Compiler-generated task fibers}{%
Highlighted statements implement compiler-inserted orchestration;
unhighlighted statements express application compute logic.}
\end{figure}

After selecting the spatial plan, \mycc lowers the final \sdfg into a
task dependency graph (\tdfg). A \emph{task} is a coarse execution unit
that groups adjacent operators with the same logical PU placement and is
implemented by \myrt as a lightweight \emph{fiber}. Each task contains
one or more \emph{kernels}, which are non-yielding device work units
later dispatched by the runtime. The compiler also creates entry and
exit tasks for host-side input and completion boundaries.

At task boundaries, \mycc inserts the communication and control
operations needed to expose asynchronous execution
(\autoref{fig:fiber-code}): DMA transfers
between storage locations and PU-local buffers, address-translation and
partition requests for near-flash work, yield points that suspend the
task fiber while an asynchronous operation is pending, and
synchronization signals that release dependent tasks only after their
inputs are ready. For example, the NPU fiber
yields after initiating a flash transfer and
signals the downstream ARM task only after writing its output scores
back to SLM.

For NPU tasks over flash-resident operands, \mycc sizes tiles against the
SLM working-set budget in the target device model and introduces double
buffering. While the NPU computes on one tile, the next tile is
transferred from flash into the alternate SLM buffer, overlapping storage
movement with dense tensor computation without requiring the full operand
to reside in device memory.

For eligible near-flash tile loops, lowering generates a page-granular,
double-buffered transfer/computation pipeline. While an NFP computes on
the current flash page, the next page is transferred into the alternate
buffer; tile boundaries on the direct near-flash path are aligned to the
flash access granularity. For filtering-style NFP tasks, parallel kernels
may additionally construct local sparse fragments that are merged before
downstream consumption.

For flash-local work, \mycc records logical page sets and emits
translation and partition requests for the runtime to resolve at task
launch (\sec\ref{s:des:rt:ftl}).

\subsection{Runtime Execution}
\label{s:des:rt}

As illustrated in \autoref{fig:rt-flow}, for each host invocation,
\myrt instantiates the compiled task graph,
schedules ready work asynchronously, and dispatches kernels to physical
processor instances and flash channels.

\subsubsection{Invocation and Completion}
\label{s:des:rt:nvme}

Compilation produces a host-side handler together with device-side
binaries and a task-graph
template. The device loads the binaries, initializes the corresponding
fibers and kernels, and registers the function with \myrt, while the
host links the handler into the application.

At invocation time, arguments are prepared according to their declared
locations. \loctype{host} inputs are staged for transfer into device
memory, \loctype{slm} arguments refer to device-local objects, and
\loctype{flash} arguments are resolved to logical flash extents. The
host then submits the function request asynchronously through
\emph{io\_uring} and NVMe, allowing multiple invocations to remain in
flight without dedicating a host thread to each request. Because data
and execution requests may arrive at the device in different orders,
\myrt tags them with an invocation identifier and uses a staged request
queue to pair each I/O request with the corresponding function invocation
before activating its task graph.

Function registration and invocation map to NVMe Computational Programs
program-management and execution requests, host--SLM transfers use the
SLM Memory Read and Write commands, and host-visible completion uses the NVMe
completion queue~\cite{nvmecsdspec.web,nvmeslmspec.web}.

Each invocation contains two special tasks. The \entrynode task transfers
required \loctype{host} inputs into device memory and resolves references
to existing \loctype{slm} objects; the \exitnode task transfers host-visible
outputs back and completes the request. Together they give \sys a
function-granularity request/response model: the host observes an
invocation's results only after the \exitnode finishes and the
corresponding NVMe completion is posted. NVMe completion is therefore
the host-visible synchronization boundary; intermediate task execution
and device-local communication remain internal to the CSD.
Semantics for SharedState are described separately in \sec\ref{s:des:state}.

\begin{figure}[t]
\centering
\includegraphics[width=\columnwidth]{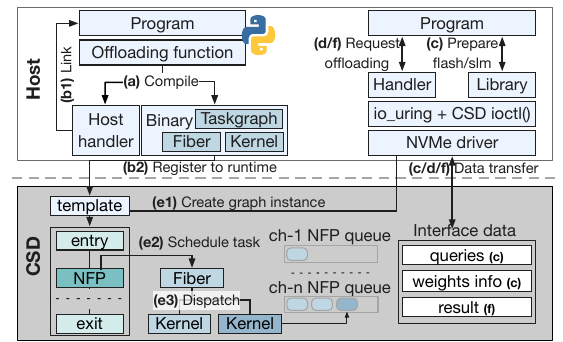}
\vminten
\mycaption{fig:rt-flow}{Runtime execution}{%
\myrt instantiates the compiled task graph for each host
invocation and schedules its tasks and kernels on the CSD.}
\end{figure}

\subsubsection{Task and Kernel Scheduling}
\label{s:des:rt:sched}

When a function request becomes executable, \myrt instantiates its
compiled task dependency graph (\autoref{fig:rt-flow}, e1) with
per-invocation metadata such as unresolved dependency counts and
kernel-dispatch status. A global
scheduler maintains ready tasks and resumes each task's fiber once its
predecessors complete (e2). Compiler-inserted yield points allow the fiber to
suspend while an asynchronous transfer or kernel is in flight; once
dispatched, a kernel runs to completion.

Task scheduling and kernel dispatch form two distinct levels. Each
compiler-assigned PU class has a kernel queue whose dispatchers select
available physical instances within that class (e3).
Independent kernels can execute concurrently across
multiple ARM cores, NFPs, or other instances of the same assigned class.
A task completes only after all its kernels finish; downstream
dependency counts are then decremented. The \exitnode task is
activated once all predecessors complete.

\myrt interleaves ready function instances so that a yielded or
resource-blocked fiber does not prevent other ready work from executing.
With interleaving disabled, it uses function-level first-come, first-served
(FCFS) scheduling.

\subsubsection{FTL-Aware Dispatch}
\label{s:des:rt:ftl}

Physical flash placement is resolved differently for shared processors
and per-channel near-flash compute. For ARM and NPU kernels, the fiber
issues ordinary flash-read requests through the FTL. The FTL performs
logical-to-physical (L2P) translation, fetches the referenced pages, and
stages the data in \loctype{slm} before computation begins.

For near-flash reads, \myrt asks the FTL at task launch to translate the
compiler-recorded logical page set to physical pages and partitions them by
their home flash channel. Each partition is then dispatched to the NFP
attached to that channel, so each NFP processes only pages resident on
its own channel.

Direct near-flash execution requires translated physical mappings to remain
valid until the offload completes. \myrt coordinates offload execution and
garbage collection (GC) with a \emph{mapping lease}. Before relocating
affected pages, GC waits for active offloads using those pages to complete.
New offloads accessing those pages wait for relocation to finish.

\subsection{Device-Resident State}
\label{s:des:state}

\myrt manages reclamation and synchronization for the CallState and
SharedState objects introduced in \sec\ref{s:des:dsl}.

\subsubsection{State Scopes}

\textit{CallState.}
CallState may live across storage accesses within a task or across
dependent tasks (fibers) of the same invocation. For example, B+ tree
traversal state in an in-storage KV lookup can live across successive
flash reads rather than be returned to the host after each access. In EC,
the comparison task on ARM produces a candidate mask that remains
device-resident until the fused precise-scoring NFP task consumes it.
No explicit programmer action is required.

\textit{SharedState.}
Objects that must outlive a complete function invocation are kept with
the \ts{epic.keep} primitive and returned as handles that later
invocations can reuse. For example, filtered select results can be kept
in SLM after independent select invocations and passed by handle to a
later join invocation. Application-managed embedding caches or model
state can likewise remain in SLM across calls until explicitly released
by the host.

\subsubsection{Lifetime and Synchronization}

\myrt allocates and reclaims device memory according to dataflow-derived
lifetimes and the \ts{epic.keep} primitive. CallState objects are
reference-counted by their consuming tasks and reclaimed when the count
reaches zero; an object consumed by a single task is therefore reclaimed
when that task completes. SharedState objects remain allocated until the
host explicitly releases them. The runtime also accounts for aggregate SLM consumption
against the SLM budget in the target device model
(\autoref{tab:sys-config}). An allocation that would exceed the budget
returns a recoverable error, allowing the application to keep the object
flash-resident instead.

Mutable device-resident objects require additional synchronization. When
a parameter is declared \texttt{mut}, \sys associates an exclusive lock
with that object. The lock is held for the critical section that accesses
the object and released when that section completes. Conflicting accesses
to the same object wait for the lock, both within and across invocations;
independent work can proceed concurrently.

Consistency between fiber tasks is enforced through the compiled task
dependency graph. Each \tdfg edge acts as a synchronization point: a
consumer is released only after its producer completes, including any
compiler-generated transfer of the producer's result into
\loctype{slm}. PUs do not require a shared hardware cache-coherence
protocol. ARM code accesses device DRAM explicitly, while NPU and NFP
kernels DMA operands between SLM and their local buffers and DMA results
back before dependent tasks are released. Task ordering and explicit data
movement thus make results visible to dependent tasks.

\subsection{Development Effort}
\label{s:des:effort}

Application developers write offloaded functions in \mydsl with input
types, shapes, and residency; the JIT frontend can infer residency.
They use \ts{epic.keep} for cross-invocation reuse, mark mutable state
with \texttt{mut}, and release retained handles when no longer needed.
The host application compiles and invokes these functions through the
device interface. Developers can focus on application logic while leaving
PU placement, fusion, data movement, and scheduling to \sys.
\sec\ref{s:eval:sota} reports application-side code size.

System vendors can build reusable hardware support that \sys automatically
uses across applications. This involves adapting code generation for the
target PUs and connecting the runtime to device memory, data transfer,
kernel dispatch, and FTL services.

\subsection{Scope and Limitations}
\label{s:dis}

\sys currently targets regular, read-heavy analytic workflows, where
large flash-resident inputs can be processed through predictable access
patterns and reused device-local state. Irregular workloads such as graph
traversals require sparse-input support beyond the current dense-input
model, while write-intensive workloads would benefit from write buffering,
coalescing, and corresponding consistency mechanisms that \sys does not
yet optimize. Processor-class mapping is static after compilation:
\mycc selects the logical PU class, and \myrt only chooses among available
instances of that class at runtime; adaptive remapping across PU classes is
future work. CSD virtualization~\cite{csdvirt.atc21}, cross-tenant isolation,
QoS-aware scheduling, and optimized scheduling across multiple CSDs
are outside the current scope.

\section{Implementation}
\label{s:impl}

\sys consists of a compiler (\mycc), runtime (\myrt), and
full-system evaluation infrastructure.
The implementation contains approximately 35K lines of C++ and Python.
We integrate \sys with FEMU~\cite{femu.fast18}, an SSD emulator widely
used for CSD research~\cite{omnicache.fast24,cemu.asplos26,sode.fast25,
dedup-ssd.asplos24,zns-compress.hpca25} and validated against real
devices~\cite{cylon.fast26,warp.fast26,cemu.asplos26}.

\mysub{Compiler}
\mycc is built on MLIR~\cite{mlir.cgo21} and implements the analyses and transformations
described in \sec\ref{s:des:cc}.
NFP kernels are lowered through the RISC-V toolchain, while NPU kernels
use the ONNX/TVM~\cite{tvm.osdi18} toolchain.
\mycc also generates host/device interfaces to register and invoke functions.

\mysub{Runtime}
\myrt handles program registration, task scheduling, and kernel dispatch
within FEMU's firmware execution threads.
It executes compiled tasks as lightweight fibers and maintains task and
object metadata.

\mysub{Evaluation platform and validation}
The host application and OS storage stack run in a KVM-accelerated VM,
while separate FEMU threads emulate NVMe processing, the FTL, and flash,
DRAM, and PCIe timing. Firmware threads are pinned to CPU cores at reduced
frequency to approximate the processing capacity of an SSD controller.

NFP and NPU kernels are simulated with Gem5~\cite{gem5.web} and
ONNXim~\cite{onnxim.cal24}, respectively.
Their per-kernel traces record cycle latencies and I/O access patterns
and are replayed at kernel dispatch in FEMU.
The traces supply accelerator timing, while runtime scheduling and
bandwidth throttling model contention for flash channels, DRAM, and PCIe.

We cross-validate Gem5 NFP timing against the Ara2 RTL~\cite{ara2.tc24}
for element-wise, reduction, and scan kernels, and ONNXim NPU timing
against the Gemmini RTL~\cite{gemmini.dac21} for matrix multiplication
and convolution.
As an end-to-end storage-system cross-check, an extended
MQSim~\cite{mqsim.fast18} using the same compute models agrees with
FEMU within 5\% on execution time for a representative subset of
workloads.

\section{Evaluation}
\label{s:eval}

Our evaluation asks four questions: whether the complete
\sys stack provides broad end-to-end benefit; how logical
mapping, fusion, and compiler-generated I/O--compute overlap contribute
to that benefit; how CallState and SharedState affect stateful workloads;
and whether the compiler cost model is accurate and the runtime machinery
is inexpensive enough to support these decisions.

\subsection{Experimental Setup}
\begin{figure*}[t!]
    \centering
    \includegraphics{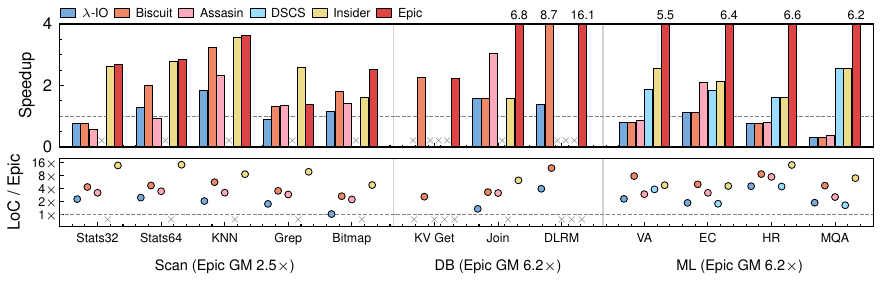}
    \vminten
    \mycaption{fig:prior-comparison}{End-to-end performance and application-side code}{%
    \textbf{Top:} throughput speedup of \cEpic and restricted
    prior-derived configurations over the respective \cHost baseline.
    \cHost is the CPU baseline for Scan/DB and TPUv4 for ML.
    \textbf{Bottom:} Application-side LoC relative to \cEpic for the same workload; each marker represents a rewrite against the corresponding prior system's programming interface.
    In both panels, $\times$ denotes an unsupported workload.}
    \vminten
\end{figure*}

\mysub{Testbed and baselines}
We evaluate the complete \sys stack using the full-system infrastructure
described in \sec\ref{s:impl}.
All experiments run on a dual-socket Intel Xeon Gold 5416S server;
the VM and emulated CSD occupy separate NUMA nodes.
The host--device link is limited to 5\,GB/s and CSD DRAM bandwidth to
25.6\,GB/s, consistent with recent CSD prototypes~\cite{smartssd-gen2.fms22,activepy.dac23}.
Unless stated otherwise, the reference CSD uses eight flash channels
(\autoref{tab:sys-config}).
ML workloads use the 32-channel configuration in \autoref{tab:sys-config}.
Throughout the evaluation, \cHost denotes the workload's host-side
baseline: the host CPU for scan and database workloads and TPUv4 for ML.
\cEpic always denotes the full configuration; \ts{NoX} names a
single-feature ablation and \ts{X-Only} a mapping restriction.
We compare \cEpic with five prior ISC systems using the configurations
summarized in \autoref{tab:sota-config}.
These state-of-the-art CSDs span different
architectures and execution models.

\begin{table}[t]
\centering
\footnotesize
\setlength{\tabcolsep}{3pt}
\begin{tabularx}{\columnwidth}{@{}l|l|>{\raggedright\arraybackslash}X@{}}
\textbf{Configuration} & \textbf{PU scope} & \textbf{Execution model} \\
\hline
\cLambdaIO~\cite{lambda-io.fast23}
    & ARM
    & Per-read, stateless; sequential I/O \\
\cBiscuit~\cite{biscuit.isca16}
    & ARM
    & Chained SSDlets; I/O--compute overlap \\
\cAssasin~\cite{assasin.micro22}
    & NFP
    & Near-flash streaming; flash interleaving \\
\cDSCS~\cite{dscs.asplos24}
    & NPU + ARM ctl.
    & Flash $\rightarrow$ DRAM $\rightarrow$ NPU execution \\
\cInsider~\cite{insider.atc19}
    & FPGA$^\dagger$
    & Streaming kernel chains \\
\rowcolor{epictint}[0pt][0pt]
\textbf{\cEpic}
    & ARM + NFP + NPU
    & Workflow mapping/fusion; overlap; state \\
\end{tabularx}
\vspace{2pt}
\mycaption{tab:sota-config}{Restricted prior-derived configurations}{%
Except for \cInsider, each configuration uses a subset of the same
reference-CSD PUs and retains the corresponding design's execution model.
$^\dagger$\cInsider uses an FPGA coprocessor whose kernel throughput
is estimated via HLS~\cite{vivado-hls.web}.}
\vminfive
\end{table}

\begin{table}[t]
\centering
\footnotesize
\setlength{\tabcolsep}{4pt}
\begin{tabular}{lp{5.8cm}}
\toprule
\multicolumn{2}{c}{\textbf{Reference Evaluation Configuration}} \\
\midrule
Host (VM) & 8 cores, 3.0\,GHz; 32\,GB DRAM; 5\,GB/s PCIe link \\
CSD ARMs & 8$\times$1.8\,GHz~\cite{cortex-r82.web}; FP64; 2\,GB DRAM, 25.6\,GB/s \\
CSD Flash  & 256\,GB, 20\,$\mu$s read, 8 ch$\times$8 chips~\cite{ares.micro24}$^\dagger$ \\
NFP ($\times$8) & 800\,MHz, FP16/32; 12.1\,mm$^2$ total, 1.3\,W total \\
               & RVV 64-wide, 8\,KB L1, 128\,KB L2 \\
NPU        & 1\,GHz, FP16$\cdot$FP32, 13.3\,mm$^2$, 6.7\,W \\
               & 4 systolic tiles (32${\times}$32 each), 8\,MB shared buffer \\
\bottomrule
\end{tabular}
\mycaption{tab:sys-config}{Reference evaluation CSD configuration}{%
$^\dagger$ML workloads use a 32-channel, enterprise-grade
configuration (32 NFPs, 48.4\,mm$^2$, 5.2\,W total NFP) with TPUv4 as
the \cHost baseline.}
\vminfive
\end{table}

\mysub{Reference-CSD feasibility}
NFP area and power follow Ara2~\cite{ara2.tc24}, and NPU estimates follow
DSCS~\cite{dscs.asplos24}.
We validate both component estimates with Synopsys Design
Compiler~\cite{synopsys-dc.web} and FreePDK45~\cite{freepdk45.web},
estimate SRAM with CACTI~\cite{cacti7.taco17}, and scale to 28\,nm
using DeepScaleTool~\cite{deepscale.iscas21}, following the methodology
of DSCS~\cite{dscs.asplos24}.
The NFP runs at 800\,MHz at 28\,nm.
The NFPs and NPU together consume 8.0\,W in the default eight-channel
configuration (1.3\,W for eight NFPs plus 6.7\,W for the NPU) and
11.9\,W in the 32-channel ML configuration.
These totals leave headroom within the 25\,W envelope of commercial
SSDs~\cite{ssd-power.web,smartssd-gen2.fms22} for the controller, DRAM,
and flash I/O.

\mysub{Workloads}
We use 12 workloads spanning three application groups.
The five scan workloads follow prior ISC
work~\cite{insider.atc19,lambda-io.fast23,assasin.micro22}:
\wStatsThirtyTwo/\wStatsSixtyFour compute the minimum, maximum, and average of
INT32/INT64 sequences; \wKNN computes Manhattan distances;
\wGrep matches rows against a target string; and \wBitmap
applies run-length encoding.
Each processes 128\,GB of synthetic data in 1\,MB batches.
The DB group contains \wKVGet, \wJoin, and
\wDLRM embedding lookup; their storage-specific execution patterns
are described with the stateful experiments in \sec\ref{s:eval:state}.
The four ML workflows are video analytics (\wVA), extreme
classification (\wEC), hybrid recommendation (\wHR), and
multi-query attention (\wMQA); their stages, storage locations, and
compiler-selected PU classes are detailed in \autoref{tab:ml_bench}.

\subsection{End-to-End Performance}
\label{s:eval:sota}
\begin{table*}[tbh]
    \footnotesize
\centering
\setlength{\tabcolsep}{4pt}
\newcommand{\var}[1]{\textit{#1}}
\newcommand{\workload}[1] {\textbf{#1}}
\newcommand{\steps}[1] {\parbox{0.44\textwidth}{\vspace{0.2em}#1\vspace{0.2em}}}
\newcommand{\param}[1] {\parbox{0.34\textwidth}{\vspace{0.2em}#1\vspace{0.2em}}}
\newcommand{\mapping}[1] {\parbox{0.13\textwidth}{\vspace{0.2em}#1\vspace{0.2em}}}
\resizebox{\textwidth}{!}{%
\begin{tabular}{l|l|l|l}
    \hline
    \textbf{Workflow} & \textbf{Procedure} & \textbf{Parameter Location, Type, and Size} & \textbf{PU Mapping} \\
    \hline
    \workload{VA} & \steps{
        1. \textbf{Calculate mean square diff.}: $v_1 \Leftarrow \frac{1}{d^2}\sum(\operatorname{grayscale}(\var{frame}) - \var{ref})^2$ \\
        2. \textbf{If $v_1\!\le\!\sigma_1$, run proxy model}: $v_2 \Leftarrow \operatorname{CNN_{\var{proxy}}}(\var{frame})$ \\
        3. \textbf{If $v_2\!\le\!\sigma_2$, update} $\var{ref} \Leftarrow \operatorname{grayscale}(\var{frame})$ and return $\var{frame}$} &
        \param{$\var{frame}:Flash,Int8[d,d,c_0], d{=}240, c_0{=}3$ \\
            $\var{ref}:SLM,Int8[d,d], batch{=}32$ \\
            $CNN_{\var{proxy}}:SLM,FP16, 245KB$} &
        \mapping{mse: nfp \\ \textbf{proxy model}: npu} \\
    \hline
    \workload{EC} & \steps{
        1. \textbf{Project query batch} to lower dimensions: $Q^{\prime} \Leftarrow Q\times P$ \\
        2. \textbf{Compute approximate scores}: $S^{\prime} \Leftarrow Q^{\prime} \times W^{\prime}$ \\
        3. \textbf{Select candidates} per query: $C_q \Leftarrow \{j\in \{1,\ldots,l\}\mid S^{\prime}_{q,j}>\sigma\}$ \\
        4. \textbf{Compute precise scores} per query: $S_{q,j} \Leftarrow Q_{q} \times W_{j}$, for $j \in C_q$} &
        \param{$Q:Host,FP16[b,d], P:SLM,FP16[d,k]$ \\
        $W:Flash,FP16[d,l], W':Flash,FP16[k,l]$ \\
        $b{=}128, d{=}1024, k{=}128, l{=}40M$} &
        \mapping{projection: npu \\ approx. score: npu \\ threshold filter: arm \\ \textbf{precise score}: nfp} \\
    \hline
    \workload{HR} & \steps{
        1. \textbf{Filter content by metadata}: $C \Leftarrow \{i \in I | \mathrm{Rules}(\var{Meta}_i)\}$ \\
        2. \textbf{Compute ranking score} per candidate: $S_{c} \Leftarrow q \times W_c$, for $c \in C$} &
        \param{$q:Host,FP32[d], W:Flash,FP16[l,d]$ \\
               $\var{Meta}:Flash,Int32[l,k], d{=}512, k{=}64, l{=}1M$} &
        \mapping{rule-based filter: nfp \\ \textbf{score calc.}: nfp} \\
    \hline
    \workload{MQA} & \steps{
        1. \textbf{Get attention weights} per head, $A_{h} \Leftarrow \operatorname{softmax}(Q_{h} \times K^{\intercal}/\sqrt{d_k})$ \\
        2. \textbf{Get weighted sum of values} per head, $O_{h} \Leftarrow A_{h} \times V$} &
        \param{$K:Flash,FP16[l,d], V:Flash,FP16[l,d]$ \\
        $Q:Host,FP16[h,d], d{=}1024, h{=}4, l{=}1M$} &
        \mapping{attention: nfp \\ \textbf{weighted sum}: nfp} \\
    \hline
\end{tabular}}
\mycaption{tab:ml_bench}{ML workflow evaluation}{%
Representative multi-stage ML workflows used to evaluate spatial and
temporal optimization. The table reports each workflow's major
operations, parameter residency/type/size, and \sys's compiler-selected
PU mapping.}
\vminfifteen
\end{table*}

We first ask whether exposing the complete located, lifetime-aware
workflow translates into broad end-to-end benefit.
\autoref{fig:prior-comparison} compares \cEpic with five
restricted prior-derived configurations across all 12 workloads.
Each is a composite design point that retains the corresponding
system's PU scope and execution model; \sec\ref{s:eval:ml} and
\sec\ref{s:eval:state} then isolate \sys's spatial, temporal, and
stateful mechanisms on fixed hardware.

\autoref{fig:prior-comparison} shows that \cEpic accelerates all
12 workloads over their respective \cHost baselines, achieving a
4.2$\times$ geometric-mean speedup overall and reaching 16.1$\times$
on \wDLRM embedding lookup.
Across workloads, \cEpic is 1.6$\times$ faster in geometric mean
than the strongest restricted prior-derived configuration for each
workload.

The restricted configurations are competitive when a workload closely
matches their execution model.
\cInsider's FPGA-like streaming path is competitive on the scan
workloads and leads on \wGrep, reaching 2.60$\times$ versus
\cEpic's 1.37$\times$.
\cBiscuit likewise nearly matches \cEpic on \wKVGet, where
executing the complete traversal on the device captures most of the
benefit and is supported by its ARM dataflow pipeline.

The gap widens when workflows require richer state or coordinated
heterogeneous stages.
Under the compiler-selected \wJoin placement used here, the selects
execute on the NFPs and \cEpic reaches 6.79$\times$ versus
3.05$\times$ for \cAssasin; the stateful experiment in
\sec\ref{s:eval:state} instead fixes the selects on ARM to isolate
SharedState.
The ML workloads likewise benefit from coordinating placement and
fusion across heterogeneous execution resources, as examined next in
\sec\ref{s:eval:ml}.

The lower panel of \autoref{fig:prior-comparison} reports
application-side lines of code.
For each prior-derived configuration, we rewrite the same workload against
that system's described programming interface and count its
application-side lines of code.
\sys uses one \mydsl program per workload, which \mycc maps across
ARM, NFP, and NPU.
In our implementations, this reduces application-side LoC by up to
14$\times$ while covering all 12 workloads with the same
compiler/runtime stack.
The restricted configurations expose narrower programming interfaces
or workload scopes, which also produces the unsupported cases marked
by $\times$ in the figure.

\subsection{Placement, Fusion, and Overlap}
\label{s:eval:ml}

\begin{figure}[t]
    \centering
    \begin{subfigure}{\columnwidth}
        \centering
        \includegraphics{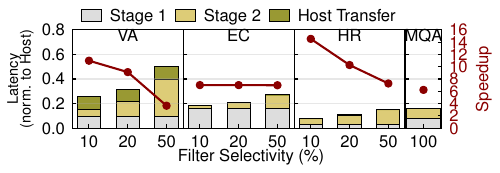}
        \vminten\vminten
        \caption{Stage-latency breakdown under a fixed compiler-selected plan.}
        \label{fig:mlbench-speedup-and-breakdown}
    \end{subfigure}
    \par
    \hspace*{-8pt}
    \begin{subfigure}[b]{0.22\textwidth}
        \centering
        \includegraphics{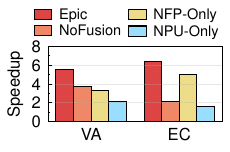}
        \vminten\vminten
        \caption{Logical mapping and fusion restrictions.}
        \label{fig:mlbench-arch}
    \end{subfigure}
    \hspace*{8pt}
    \begin{subfigure}[b]{0.22\textwidth}
        \centering
        \includegraphics{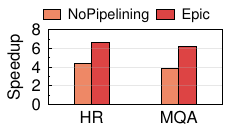}
        \vminten\vminten
        \caption{Software pipelining for HR and MQA.}
        \label{fig:mlbench-nfp-config}
    \end{subfigure}
    \vminfive
    \mycaption{fig:mlbench}{Spatial and temporal workflow optimization}{%
    \textbf{(a)} Latency normalized to \cHost; the red line
    (right axis) shows speedup over \cHost.
    The compiler-selected plan is fixed while filter selectivity varies
    for \wVA, \wEC, and \wHR; \wMQA is the non-filtering
    contrast.
    \textbf{(b)} Throughput speedup on \wVA and \wEC with fusion
    disabled (\cNoFusion) or mapping restricted to a single PU class
    (\ts{X-Only}), compared with the full \cEpic plan.
    \textbf{(c)} Throughput speedup for \cEpic vs.\ \cNoPipelining
    (NFP kernel pipelining disabled) on \wHR and \wMQA.}
\end{figure}

\mysub{Stage breakdown}
\autoref{tab:ml_bench} summarizes the four ML workflows used to
evaluate placement and data-movement effects across stages.
\wVA detects scene changes by comparing flash-resident frames with
a reference and conditionally running a small proxy
model~\cite{noscope.vldb17,tahoma.icde19,smol.vldb20}.
\wEC uses approximate screening before precise scoring over a much
larger class set~\cite{ecssd.isca23,ec.micro21}.
\wHR filters candidates using flash-resident metadata before
ranking them~\cite{hr.umuai02}.
\wMQA computes attention weights and then weighted sums over
flash-resident key/value matrices~\cite{mqa.arxiv19}.

\autoref{fig:mlbench-speedup-and-breakdown} breaks \cEpic's latency
into \cStageOne, \cStageTwo, and \cHostTransfer, normalized to
\cHost.
For \wVA, \wEC, and \wHR, we sweep filter selectivity while
holding the compiler-selected plan fixed; selectivity changes how much
data and work reach the downstream stage, not the mapping itself.
\wMQA is the non-filtering contrast and is shown at its fixed
operating point.

The resulting stage balance differs sharply across workflows.
As selectivity rises, more data reaches the downstream stage in
\wVA and \wHR, so that stage becomes the bottleneck; for
\wVA, speedup consequently falls from 10.0$\times$ at 10\%
selectivity to 3.3$\times$ at 50\%.
In contrast, \wEC remains dominated by its first stage and sustains
6.4$\times$ speedup across the sweep, while the non-filtering
\wMQA has comparatively balanced stages.
These different bottleneck structures motivate coordinating PU mapping
and fusion over the workflow rather than fixing one execution class for
all stages.

\begin{figure*}[t!]
    \centering
    \includegraphics[width=.90\textwidth]{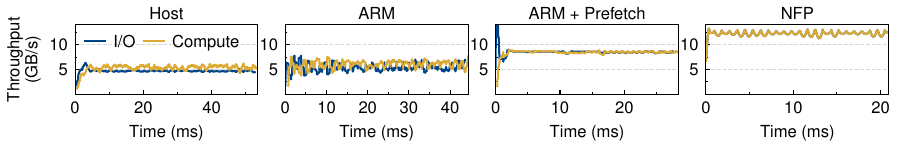}
    \vminfive
    \mycaption{fig:microbenchmarks}{I/O--compute overlap in Stats64}{%
    Time series of I/O and compute throughput for \wStatsSixtyFour under
    \cHost, \cARM, \cARMPrefetch, and \cNFP.
    \cARMPrefetch enables compiler-generated tiling and prefetching;
    \cNFP executes on channel-local processors.}
    \vminten
\end{figure*}

\mysub{Placement and fusion}
\autoref{fig:mlbench-arch} compares the full \cEpic plan for
\wVA and \wEC with \cNoFusion, \cNFPOnly, and
\cNPUOnly.
All four configurations compile the same \mydsl program; compiler options
either restrict the eligible PU set or disable fusion while retaining
the rest of the compilation flow.
\cEpic leads all three restrictions on both workloads because the
stages benefit from different execution resources and from eliminating
movement at their boundaries.

\wEC makes this interaction especially clear: \cEpic reaches
6.36$\times$, while \cNoFusion falls to 2.13$\times$, below the
4.99$\times$ \cNFPOnly result.
Without fusion, candidate selection materializes the selected weights in
SLM before precise scoring. Fusing selection with the NFP precise-scoring
kernel avoids this intermediate while keeping other stages on suitable PUs.

\mysub{I/O--compute overlap}
The \wStatsSixtyFour timeline in
\autoref{fig:microbenchmarks} shows I/O--compute overlap from the
compiler's tiling and prefetching on ARM.
\cHost is constrained by the host I/O path, while \cARM
without prefetching exposes periodic stalls between flash I/O and
computation.
\cARMPrefetch fetches the next tile while the ARM cores compute
on the current one.
\cNFP instead moves the work onto channel-local near-flash
processors and further shortens the execution interval.

\autoref{fig:mlbench-nfp-config} evaluates compiler-generated
software pipelining inside the NFP kernel.
For \wHR and \wMQA, both configurations use the same RVV-64 NFP;
\cNoPipelining disables compiler-generated NFP software pipelining,
while \cEpic enables it.
Pipelining improves throughput by 1.50$\times$ on \wHR and
1.59$\times$ on \wMQA by overlapping page movement with computation
inside the near-flash execution path.
Thus, ARM prefetching and NFP software pipelining both create
I/O--compute overlap, but at different levels of execution.

\subsection{Stateful Execution}
\label{s:eval:state}
\begin{figure*}[t]
    \begin{subfigure}[b]{0.25\textwidth}
        \centering
        \includegraphics[width=\textwidth]{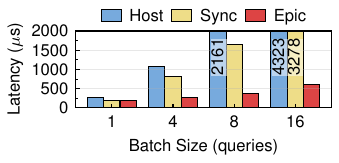}
        \caption{KV Get latency vs.\ batch size.}
        \label{fig:kvget-latency}
    \end{subfigure}%
    \begin{subfigure}[b]{0.25\textwidth}
        \centering
        \includegraphics[width=\textwidth]{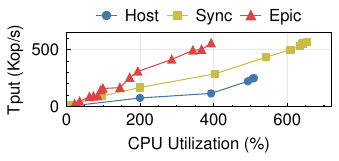}
        \caption{KV Get Tput vs.\ CPU util.}
        \label{fig:btree-pareto-cpuutil}
    \end{subfigure}%
    \begin{subfigure}[b]{0.25\textwidth}
        \centering
        \includegraphics[width=\textwidth]{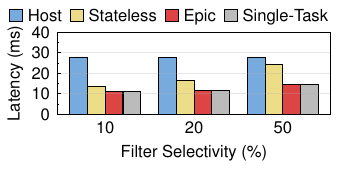}
        \caption{Join latency vs.\ filter selectivity.}
        \label{fig:relation-join}
    \end{subfigure}%
    \begin{subfigure}[b]{0.25\textwidth}
        \centering
        \includegraphics[width=\textwidth]{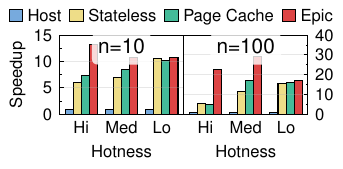}
        \caption{DLRM lookup speedup.}
        \label{fig:dlrm-speedup}
    \end{subfigure}
    \vmintwenty
    \mycaption{fig:db-eval}{Stateful execution across access and invocation boundaries}{%
    \textbf{(a--b) KV Get:}
    \cHost is the host-side baseline; \cSync executes the complete
    B+ tree traversal on the device, where its state lives across
    successive flash accesses, but submits invocations synchronously;
    \cEpic uses the same device-side traversal semantics with
    asynchronous submission to maintain many concurrent traversals.
    \textbf{(c) Join:}
    \cHost runs the query on the host; \cStateless offloads the
    selects and performs the join on the host; \cEpic issues separate
    select and join offloads, with filtered select results that live on
    the device across invocations; \cSingleTask executes the three
    operators as one offloaded task and serves as a reference.
    \textbf{(d) DLRM embedding lookup:}
    \cHost is the no-offload baseline; \cStateless performs device
    lookups without application-managed state that outlives an invocation;
    \cPageCache adds a generic shared page cache; and \cEpic uses
    an application-managed vector cache that outlives invocations via
    \ts{epic.keep}. The left and right halves use $n{=}10$ and $n{=}100$,
    respectively, with separate y-axes.}
    \vminten
\end{figure*}

We evaluate CallState across storage accesses with \wKVGet
and SharedState with \wJoin and \wDLRM embedding lookup. The
multi-stage workflows in \sec\ref{s:eval:ml} also use CallState
intermediates passed between dependent tasks within an invocation.
In these experiments, all device-side operators execute on the ARM
cores, so the comparisons vary state lifetime and submission rather
than placement.
The B+ tree traversal in \wKVGet and cache management in
\wDLRM are control-intensive and target ARM; for \wJoin, we
likewise fix the select operators on ARM even though they are eligible
for NFP execution.
The end-to-end \wJoin result in \autoref{fig:prior-comparison}
instead uses the compiler-selected placement, which maps the selects to
the NFPs.
These experiments therefore isolate the lifetime boundaries that
materially change their storage execution paths.

\mysub{CallState: KV Get}
A flash-resident B+ tree lookup follows multiple dependent page reads
to traverse the index and retrieve a value, making repeated host
mediation expensive; prior in-storage KV systems therefore offload the
complete traversal~\cite{sode.fast25}.
We evaluate \wKVGet on a 68\,GB flash-resident B+ tree with five
index layers, following prior setups~\cite{sode.fast25,xrp.osdi22};
the first three index layers are cached in host memory and the leaf
layer stores 64\,B values.
\cHost performs the traversal through the host-side storage path.
\cSync offloads the complete B+ tree traversal to the CSD and uses
synchronous submission, so the traversal state lives on the device
across successive flash accesses instead of returning control and state
to the host between reads.
\cEpic uses the same device-side traversal semantics but submits
invocations asynchronously, allowing the runtime to maintain and
interleave many independent stateful traversals while their flash reads
are outstanding.

In the latency-oriented experiment, one worker thread per core issues
batches of queries; we sweep batch size and measure batch completion
latency (\autoref{fig:kvget-latency}).
At batch size 1, \cSync and \cEpic have the same latency, both below
\cHost, reflecting the common benefit of keeping the complete traversal
and its state on the device.
As concurrency grows, asynchronous submission separates them:
at batch size 16, \cEpic completes a batch in 614\,$\mu$s on average,
versus 3278\,$\mu$s for \cSync, because multiple traversal fibers can
remain in flight while individual flash reads are outstanding.

In the throughput-oriented experiment, we scale host worker threads
for \cHost and \cSync, and queue depth for \cEpic.
\autoref{fig:btree-pareto-cpuutil} shows that \cEpic reaches
essentially the same peak throughput as \cSync, but uses 393\%
host CPU versus 653\% for \cSync.
Asynchronous submission therefore converts host-thread concurrency into
device-side concurrency: traversal fibers retain their independent state,
yield on flash reads, and allow other traversals to make progress.
With a single host thread, increasing the number of in-flight
\cEpic requests from 2 to 8 achieves 91\% of ideal linear throughput
scaling while adding under 10\% latency.

\mysub{SharedState: Join}
Relational queries compose multiple operators, so independently invoked
offloads benefit when one operator's output can remain device-resident
for a later operator rather than returning through the host.
We evaluate
\ts{join(sel(table1), sel(table2))} on two 512k-row tables with
128\,B rows, sweeping a common filter selectivity~$p$
(\autoref{fig:relation-join}).
The two select operators and the join are issued as separate offloads.
\cHost performs the complete query on the host.
\cStateless offloads the two selects but returns their filtered
results so the host performs the join.
\cEpic invokes the two selects and join separately; the filtered
select results live on the device across the invocation boundary until
the join consumes them.
\cSingleTask executes all three operators as one offloaded task and
serves as a reference.

As \autoref{fig:relation-join} shows, \cStateless benefits from
returning only filtered rows, but its cost grows as more of those rows
must cross the host boundary for the join.
\cEpic keeps the filtered intermediates on the device and returns
only the final result; at $p{=}50\%$, latency is 14.5\,ms versus
24.5\,ms for \cStateless.
\cSingleTask matches \cEpic across the sweep, showing
that exchanging the intermediate across separate invocations adds no
measurable latency at this scale.

\mysub{SharedState: DLRM embedding lookup}
Embedding lookups dominate DLRM inference latency for large
tables~\cite{dlrm.isca23,rabbitail.sac25,rects.systor24}.
CSD offloading alleviates host memory pressure and reduces data
movement by serving these lookups near
storage~\cite{ndrec.tc24,rmssd.hpca22,recssd.asplos21,screc.tc26,bam.asplos23}.
Following~\cite{dlrm.isca23}, we use 16 tables with 128-dimensional
FP32 embeddings, scale each table to $10^7$ vectors, and generate
requests at different hotness levels.
Each table has one worker thread retrieving $n$ vectors per query, and
we evaluate $n{=}10$ and $n{=}100$ (\autoref{fig:dlrm-speedup}).

\cHost performs lookups without offloading and allocates per-thread
host buffers for 1\% of the vectors.
\cStateless offloads table-wise lookups and returns compact results.
\cPageCache adds a generic shared, page-granular device cache.
\cEpic instead uses application-managed, table-private,
vector-granular FIFO caches in SLM that outlive lookup invocations via
the \ts{epic.keep} primitive.
This organization can keep hot vectors without consuming cache space
for unrelated vectors on the same flash page or causing entries from
different tables to compete in one shared cache.

\autoref{fig:dlrm-speedup} shows that \cStateless benefits from
offloading, \cPageCache adds page-level reuse, and \cEpic
further improves performance through application-managed
cross-invocation vector reuse, reaching 16.1$\times$ over \cHost.
Across the evaluated settings, \cEpic outperforms \cPageCache
by a geometric mean of 1.65$\times$.
Because their cache organizations differ,
this comparison measures the benefit of the application-managed
vector cache rather than synchronization overhead.

To isolate synchronization overhead, we additionally evaluate
\cSharedFIFO using the same vector-granular FIFO policy and cache
capacity as \cEpic, but with one cache shared by all tables behind
a single lock.
This comparison changes the synchronization organization while holding
the cache policy and capacity fixed.
Across the six $n$/hotness settings, \cSharedFIFO is
1.03--1.54$\times$ slower than \cEpic, with a geometric mean of
1.19$\times$; the largest penalties occur at $n{=}100$, where each
request performs more cache accesses under the shared lock.
\cEpic instead uses table-private state and a mutex per cache,
so accesses to different tables do not serialize
(\sec\ref{s:des:state}).

\subsection{Cost Model and Overheads}
\label{s:eval:overhead}

\mysub{Cost-model accuracy}
\mycc's movement-aware cost model drives logical mapping and
cost-directed fusion.
Across the scan workloads and ML pipelines, predicted throughput is
within a geometric-mean factor of $1.24\times$ of measured throughput,
with rank correlation $1.0$.
Because plan selection depends on relative ordering, the perfect rank
correlation supports the model's use for compiler-selected decisions
even when absolute predictions are not exact.

\mysub{System overheads}
The request path adds little latency: an NVMe ISC round trip takes
${\sim}$9\,$\mu$s from host \emph{io\_uring} submission through
completion, including ${\sim}$1.5\,$\mu$s of runtime task scheduling.
Scheduling and dispatch overlap with flash I/O, contributing less than
1\% of total workload time.

Lower-level runtime costs are smaller still: a fiber switch takes
${\sim}$120\,ns and kernel dispatch ${\sim}$80\,ns, both well below a
single flash-page read of ${\sim}$20\,$\mu$s.

Compilation and registration take ${\sim}$1\,s and 6.3\,ms,
respectively; both are one-time costs outside request execution.
The runtime occupies 2.9\,MB of device DRAM, while compiled \mydsl
binaries range from 16 to 150\,KB.

\section{Related Work}
\label{s:rel}

\subsection{ISC Systems}

Prior ISC systems expose different portions of the storage execution path.
\autoref{tab:rel} compares representative systems along
five storage-facing dimensions.
Summarizer, IceClave, and FlashAbacus execute offloaded user code as opaque
kernels, while $\lambda$-IO and ActivePy minimize application changes through
I/O interception and transparent execution, respectively~\cite{summarizer.micro17,
iceclave.micro21,flashabacus.eurosys18,lambda-io.fast23,activepy.dac23}.
IceClave provides trusted execution and isolation for concurrent in-storage
programs, addressing a complementary concern to \sys's workflow optimization.
Biscuit exposes a dataflow API, while Insider and MetalFS provide
programmable FPGA interfaces; other systems expose fixed operators or
specialize the data path for particular applications~\cite{biscuit.isca16,
insider.atc19,metalfs.eurosys20,dynamic.micro19,nds.micro21}.
\emph{Offload scope} captures how offloaded computation is structured and exposed to the system,
from intercepted I/O and opaque kernels to operator graphs, pipelines, and
storage-typed DSL programs, while \emph{device-local state} captures whether application data
is retained on the device across independent offloads.
\sys accepts storage-typed DSL programs, reconstructs the workflow
automatically, and explicitly supports SharedState through \ts{epic.keep}.

\setlength{\tabcolsep}{2.0pt}

\newcommand{\citeapp}{\cite{nascent.fpga21,
bluedbm.isca15, grafboost.isca18, extrav.vldb17, rmssd.hpca22,
recssd.asplos21, cognitivessd.atc19, holistic-gnn.fast22, glist.atc21,
smart-inf.hpca24, genstore.asplos22, inspire.isca22,
ecssd.isca23, optimstore.hpca23, deepstore.micro19, beacongnn.hpca24,
instinfer.hpca25, behemoth.fast21, yoursql.vldb16, smartsage.isca22,
fpim3d.micro22,
crossbit.micro25, sting.dac24, iskeva.lctes22, ibex.vldb14,
smartgraph.socc24, cord.ipdps25}}

\newcommand{\citefixed}{\cite{dynamic.micro19,nds.micro21}}

\newcommand{\citekernel}{\cite{summarizer.micro17,iceclave.micro21,flashabacus.eurosys18}}

\newcommand{\mrot}[1]{%
  \raisebox{0ex}{\makebox[0pt][l]{%
    \begin{rotate}{28}{\textbf{#1}}\end{rotate}}}}
\newcommand{\mcc}[1]{%
  \multicolumn{1}{c}{\mrot{{\scriptsize #1}}}}

\begin{table}[t!]
\centering
\scriptsize
\vten
\begin{tabular}{@{}llcccc@{}}
\textbf{System} & \shortstack[l]{\textbf{Offload}\\\textbf{scope}}
  & \shortstack{\textbf{Device-}\\\textbf{local state}}
  & \shortstack{\textbf{Workflow}\\\textbf{opt.}}
  & \shortstack{\textbf{Async}\\\textbf{scheduling}}
  & \shortstack{\textbf{Hetero.}\\\textbf{mapping}} \\
\hline
Summarizer et al.\textsuperscript{\S}
  & Kernel offload    &     &     &     &     \\
$\lambda$-IO~\cite{lambda-io.fast23}
  & I/O intercept     &     &     &     &     \\
ActivePy~\cite{activepy.dac23}
  & Transparent       &     &     &     &     \\
DockerSSD~\cite{dockerssd.hpca24}
  & Containers        & \hy &     &     &     \\
Biscuit~\cite{biscuit.isca16}
  & Dataflow graph    &     &     & \yes&     \\
Assasin~\cite{assasin.micro22}
  & Stream kernel     &     & \hy & \hy &     \\
DSCS~\cite{dscs.asplos24}
  & DL operators      &     &     & \hy & \hy \\
\begin{tabular}[b]{@{}l@{}}Insider~\cite{insider.atc19},\\MetalFS~\cite{metalfs.eurosys20}\end{tabular}
  & Operator pipeline &     &     & \yes& \hy \\
FixOps$^\dagger$
  & Fixed operators   &     &     &     &     \\
Apps$^\ddagger$
  & App-specific      &     & \hy &     &     \\
Conduit~\cite{conduit.hpca26}
  & Loop-level        &     & \hy &     & \yes\\
\rowcolor{epictint}[0pt][0pt]
\textbf{\sys}
  & \begin{tabular}[b]{@{}l@{}}\textbf{Storage-typed}\\\textbf{DSL}\end{tabular}
                       & \yes& \yes& \yes& \yes\\
\end{tabular}

\mycaption{tab:rel}{Representative ISC programming and execution systems}{%
Comparison along five storage-facing dimensions.
\yes~= full support (provided automatically by the system);
\hy~= partial support (realized manually by the developer or supported only
within a narrow scope); blank = unsupported.
\textsuperscript{\S} Kernel-offload systems~\citekernel.
$^\dagger$ Fixed-operator systems~\citefixed.
$^\ddagger$ Application-specific ISC systems~\citeapp.}
\vminfive
\end{table}

\emph{Workflow optimization} captures whether dependent operations are
transformed as one storage workflow rather than as isolated kernels.
Assasin provides streaming/locality optimizations, and application-specific
systems employ workload-specific transformations~\cite{assasin.micro22,
nascent.fpga21,genstore.asplos22,ecssd.isca23,optimstore.hpca23}.
DeepStore supports programmable similarity queries on application-selected
SSD- or channel-level tensor cores, with caching and flash
prefetching~\cite{deepstore.micro19}.
However, these interfaces do not combine logical placement, cross-operator
fusion, data movement, and lifetime information within a single
application-level dependency graph.
\emph{Asynchronous scheduling} captures whether independent I/O and
computation can proceed concurrently, a capability present in dataflow and
FPGA systems such as Biscuit, Insider, and MetalFS.
SODE exposes explicit APIs for selective on-device resubmission of
data-dependent reads~\cite{sode.fast25}.
\emph{Heterogeneous
mapping} asks whether the system itself maps or dispatches work across
multiple in-device resource classes. DSCS provides a domain-specific form
of heterogeneous execution~\cite{dscs.asplos24}; concurrent work
Conduit~\cite{conduit.hpca26} dispatches work at loop and instruction
granularity across ARM cores and DRAM- and flash-based PuM resources. \sys instead derives
heterogeneous placement from the application workflow and separates
compiler-selected logical PU placement from runtime physical binding.
At the multi-device level, CORD coordinates parallel query execution
across CSDs while accounting for data locality~\cite{cord.ipdps25}.
OmniCache coordinates host and device caches for concurrent I/O and data
processing, with processing buffers and model-driven
offloading~\cite{omnicache.fast24}.

\subsection{Programming Frameworks}

\sys is related to accelerator programming and optimization frameworks
such as Halide and TVM, which optimize operation graphs using fusion,
tiling, memory placement, and
cost-directed scheduling~\cite{halide.pldi13,tvm.osdi18,ansor.osdi20,
flextensor.asplos20,amos.isca22,tenet.isca21,glow.arxiv18}; HeteroCL, Allo,
and related heterogeneous compilers similarly separate high-level
computation from hardware-specific
mapping~\cite{heterocl.fpga19,allo.pldi24}. HPC systems such as
Legion/Regent and Chapel expose data locality and execution
placement~\cite{legion.sc12,regent.sc15,chapel.ijhpca07}.
G10 and Teraio use tensor lifetime information to schedule data movement
between GPU memory and storage~\cite{g10.micro23,teraio.neurips25}.

For CSDs, \sys plans across the storage path itself.
Using residency and lifetime, it jointly optimizes placement on
heterogeneous in-device PUs, data movement, I/O overlap, and reuse
to produce an end-to-end workflow plan.

\section{Conclusion}
\label{s:conc}

We presented \sys, an NVMe-based ISC stack that uses data residency and
lifetime to automatically coordinate workflow execution across
heterogeneous device resources. The compiler optimizes computation and
data movement across the workflow, while the runtime coordinates execution
and state management on the device.
Across 12 workloads, \sys achieves
speedups of 4.2$\times$ over host baselines and 1.6$\times$ over the
strongest prior system, on average. It also reduces
application-side code by up to 14$\times$ in our implementations.

\clearpage
\setlength{\bibsep}{2pt plus 0.1ex}
{

}

\end{document}